\documentstyle[preprint,aps,epsf]{revtex}
\begin{document}

\title{ The Thermal  Phase Transition in  Nuclear Multifragmentation:
The Role of  Coulomb Energy and Finite Size}
\author{B. K. Srivastava$^1$, S. Albergo$^2$, F. Bieser$^6$, F. P. Brady$^3$, 
Z. Caccia$^2$, D. A. Cebra$^3$, A. D. Chacon$^7$, 
J. L. Chance$^3$, Y. Choi$^ 1$,
S. Costa$^2$, J. B. Elliott$^1$, M. L. Gilkes$^1$, J. A. Hauger$^1$,
A. S. Hirsch$^1$, E. L. Hjort$^1$,
A. Insolia$^2$, M. Justice$^5$, D. Keane$^5$, 
J. C. Kintner$^3$, V. Lindenstruth$^4$,
M. A. Lisa$^6$, H. S. Matis$^6$, M. McMahan$^6$, C. McParland$^6$,
W. F. J. M\"{u}ller$^4$,
D. L. Olson$^6$, M. D. Partlan$^3$, N. T. Porile$^1$, R. Potenza$^2$,
G. Rai$^6$, J. Rasmussen$^6$, H. G. Ritter$^6$, J. Romanski$^2$, 
J. L. Romero$^3$, G. V. Russo$^2$,
H. Sann$^4$, R. P. Scharenberg$^1$, A. Scott$^5$, 
Y. Shao$^5$, T. J. M. Symons$^6$, M. Tincknell$^1$,
C. Tuv\'{e}$^2$, S. Wang$^5$, P. Warren$^1$, H. H. Wieman$^6$,
T. Wienold$^6$, and K. Wolf$^7$\\
(EOS Collaboration)}
\address{$^1$Purdue University, West Lafayette, IN 47907 \\
$^2$Universit\'{a} di Catania and Istituto Nazionale di Fisica
Nucleare-Sezione di Catania,\\
95129 Catania, Italy \\
$^3$University of California, Davis, CA 95616\\
$^4$GSI, D-64220 Darmstadt, Germany \\
$^5$Kent State University, Kent, OH 44242 \\
$^6$Nuclear Science Division, Lawrence Berkeley National Laboratory, 
Berkeley, CA 94720 \\
$^7$Texas A\&M University, College Station, TX  77843}
\date{\today}

\maketitle

A systematic analysis of the moments of the fragment size distribution
 has been carried out for the multifragmentation (MF)
of 1A GeV Au, La, and Kr on carbon. 
 The breakup of  Au and La is consistent with a continuous 
thermal phase transition.
The data indicate that the excitation energy per nucleon and isotopic 
temperature at the critical point
decrease with increasing system size. This trend is attributed 
primarily to the increasing Coulomb energy with finite size 
effects playing a smaller role.


\vspace{24pt}

The EOS collaboration has recently studied the multifragmentation (MF) of
1A GeV Au on carbon 
\cite{gilkes94,hauger96,elliott96,hauger98,elliott98,lauret98,sriv,hauger00,elliott00,scharen}.
One of the important results was the possible observation of critical
behavior and the extraction of associated critical exponents 
\cite{gilkes94,elliott96,elliott98}. The values of these exponents were 
very close to those
of ordinary fluids indicating that MF may arise from a continuous phase
transition and may belong to the same universality class as ordinary fluids.
Another important result was the successful description of the EOS MF data
by statistical thermodynamical
models, which describe 
quantum mechanically the MF of a charged nucleus \cite{scharen,bondorf,gross1,gross2,scharen2,srivas2}. 
In this paper we analyze the recent results for MF of 1AGeV La and
Kr on C \cite{hauger00} along with those previously reported for Au \cite{gilkes94,elliott96,elliott98}
in the manner proposed by Campi\cite{campi1,campi2,campi3}.
Our analysis provides the first experimental
evidence for the evolution of the MF mechanism with increasing projectile 
size and for the effects of Coulomb energy and finite size.

 The reverse kinematics experiments and the analysis by which the 
equilibrated remnant, which undergoes MF, was separated from promptly
emitted particles and the details of the determination of the remnant mass and excitation energy
are given in our earlier publications
\cite{hauger96,hauger98,hauger00}.

Campi \cite{campi1,campi2,campi3} and Bauer \cite{bauer2,bauer3} first 
suggested that the 
methods used in 
percolation studies may be applied to MF data.
In percolation theory the moments of the cluster distribution contain the 
signature of critical behavior \cite{stauffer}. The method of moments analysis
was used by several groups 
\cite{bauer,kreutz93,belkacem96,mastinu98} to search for evidence 
of the liquid-gas phase 
transition in MF.
Thus for each event, we determine the total multiplicity of charged fragments,
{\it m\/}, and the
number of charged fragments $n_{A}$, of nuclear charge Z and 
mass A \cite{hauger96}.
We calculate the {\it k\/} moments of the cluster size distribution given by
\begin{equation}
{M_k(m)}~ = ~ \sum A^k n_{A}(m)
\end{equation}
where the sum runs over all masses in the event  including neutrons except for the heaviest fragment. 
This quantity was instrumental in extracting  the critical exponents 
in Au+C data 
\cite{gilkes94}. It has been argued that there should be an
enhancement in the critical region of the moments, $M_{k}$, for {\it k\/} $>$ $\tau-1$,
with critical exponent $\tau>$ 2 \cite{campi1,campi2}. For example, the  reduced
variance $\gamma_{2}$, i.e. the combination of moments given by
\begin{equation}
\gamma_{2} = M_{2} M_{0}/M_{1}^2 
\end{equation} 
has a peak value of 2
for a pure exponential distribution, $n_{A}\sim$ $e^{-\alpha A}$, 
regardless of
the value of $\alpha$, but $\gamma_{2}$  $>$ 2 for a power law
distribution, $n_{A}\sim$ $A^{-\tau}$, provided the system is large
enough. 
Here $M_{1}$ and $M_{2}$ are the first and second moments of the 
mass distribution in an event and $M_{0}$ is the total multiplicity  including neutrons.
\begin{figure}
\epsfxsize=8.5cm
\centerline{\epsfbox{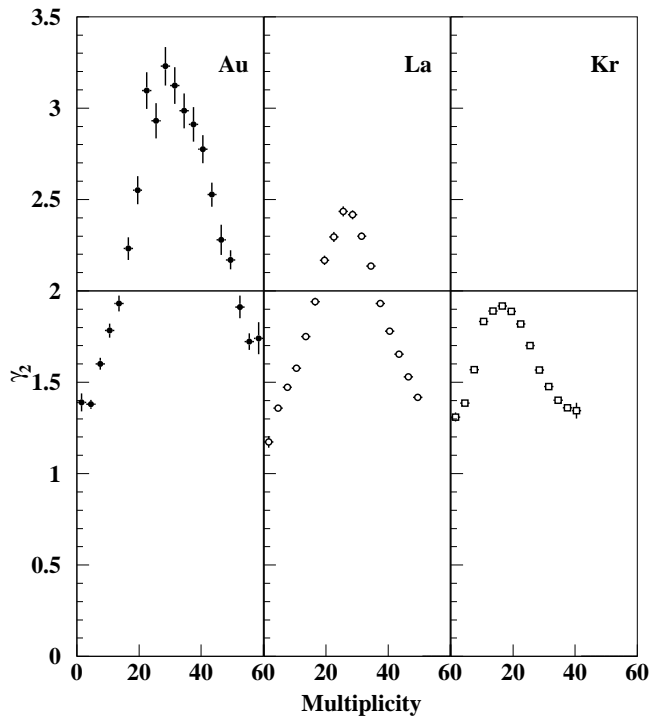}}
\vspace*{0.003in}
\caption{
$\gamma_{2}$ as a function of multiplicity from Au, La, and Kr systems
}
\label{fig.1}
\end{figure}
We have calculated  $\gamma_{2}$  event by event as a function of total
charged particle multiplicity for all three systems, as shown in 
Fig.1. It is clear that for Au and La  $\gamma_{2}>2$ at
the peak, while for Kr $\gamma_{2}<2$ . 
The position of the maximum $\gamma_{2}$ value defines the critical point,
 $m_{c}$, 
where the fluctuations in the fragment sizes are the largest. To obtain $m_{c}$ 
accurately for each system  a high resolution version of Fig.1, with points
corresponding to each value of {\it m\/}, 
 was fitted  with a polynomial of order 3-9 and the
fit with the best chi-square per degree of freedom was then chosen 
to locate the multiplicity at which $\gamma_{2}$ reaches a maximum.  
The decrease in $\gamma_{2}$ with decrease in system size observed in Fig.1 is also seen in
3d percolation studies and has been  attributed to finite size 
effects\cite{campi3}.

\begin{figure}[ht]
\epsfxsize=8.5cm
\centerline{\epsfbox{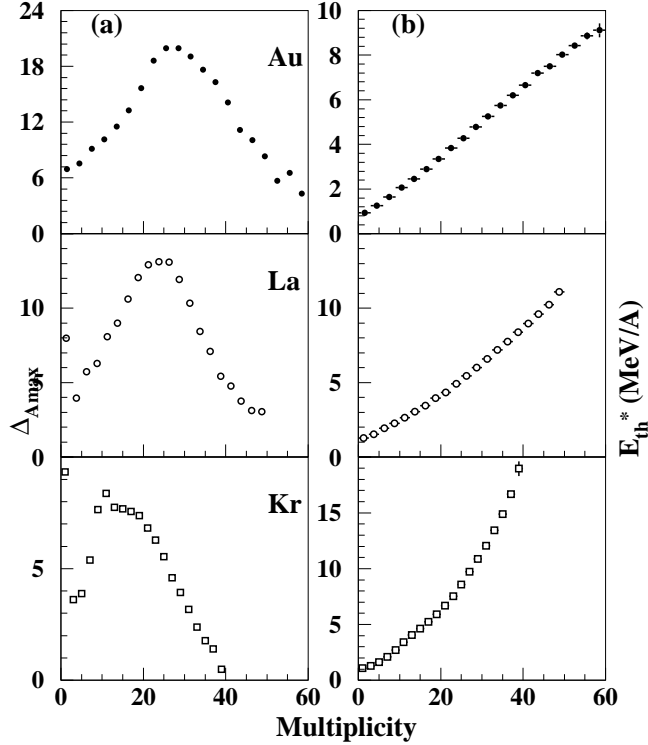}}
\caption{
 a).Fluctuations in the size of the largest fragment as a function of multiplicity.
 b).Excitation energy as a function of multiplicty. 
}
\label{fig.2}
\end{figure}

 Another way of identifying the critical point is from the fluctuations in the size of the
largest fragment. 
 The fluctuations in this quantity,  $\Delta_{Amax}$, peak at the critical point as shown in Fig.2(a).
 For Kr the peak in $\Delta_{Amax}$ is not as well defined as for Au and La. One sees a peak in $\Delta_{Amax}$ for Kr at
  {\it m\/} $\sim$10. This corresponds to 
 $\sim$ 3 MeV/nucleon excitation energy, which is too low for MF to occur. 
Thus in case of Kr the
 $m_{c}$ value was obtained from Fig.1 only.
For Au and La $m_{c}$ was obtained as the average of the two peak values in Figs.1 and 2.    
 The $m_{c}$ values for Au, La, and Kr are 
28$\pm$3, 24$\pm$3, and 18$\pm$2, respectively.
The $m_{c}$ value for Au is in 
agreement with our earlier reported values for Au within the  
respective uncertainities\cite{gilkes94,elliott98}. 

The thermal excitation energy, $E^*_{th}$, i.e. the energy available for particle 
and fragment emission, is a more fundamental quantity than the multiplicity. 
The experimental relation between these two quantities \cite{hauger00} is
shown in Fig.2(b). 
\begin{figure}[ht]
\epsfxsize=8.5cm
\centerline{\epsfbox{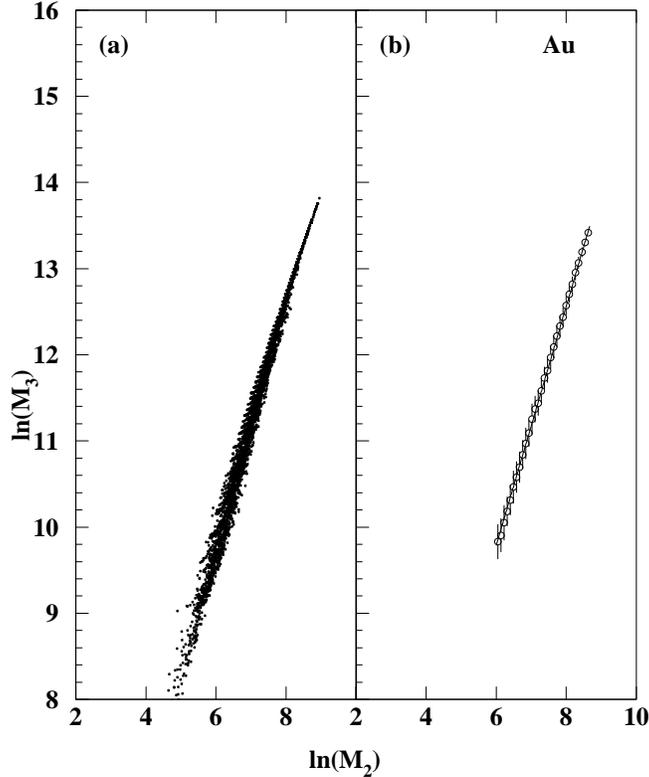}}
\caption{
a).$ln(M_{3})  vs  ln(M_{2})$ for Au above the critical multiplicity.
b). The average value of $ln(M_{3})$ at a given $ln(M_{2})$.
}
\label{fig.3}
\end{figure}  

To extract the power law exponent $\tau$, we examine the MF region above
$m_{c}$, i.e. the region past the peak in Fig.1 \cite{elliott94}.
Fig.3(a) shows a scatter plot of $ln(M_{3})$ vs $ln(M_{2})$ for Au 
\cite{gilkes94,campi1,bauer}. The slope, {\it S\/},  of the line through 
the points is related to the exponent $\tau$,  {\it S\/}=$(\tau-4)/(\tau-3$).
To obtain $\tau$ we plot the average value of $ln(M_{3})$ vs $ln(M_{2})$,
Fig.3(b). We fit the region between  $E^*_{th}$= 5.5 -7.5 MeV/nucleon 
to obtain the $\tau$ value. 
The lower energy is $\sim$ 1 MeV/nucleon 
higher than the energy corresponding to the peak in $\gamma_{2}$ and
 the higher value is close to the
 end of  $\gamma_{2}$ branch above $m_{c}$ in Fig.1.
We obtain $\tau = 2.16\pm0.08$ for Au with $\chi^2/dof$ of 1.
This value is in agreement with the $\tau$ value from
the single parameter fit, 
$n_{A}=q_{0}A^{-\tau}$, at $m_{c}$ \cite{elliott98,stauffer}.
 The same procedure was followed to fit  $ln(M_{3})$ vs $ln(M_{2})$
for La as shown in Fig.4(b), derived from the scatter plot for La in Fig.4(a). 
For La we obtain  $\tau = 2.10\pm0.06$
, with $\chi^2/dof\sim$ 6, 
again in agreement with the value obtained
from the one-parameter fit at  $m_{c}$.  

The data for Kr are shown in Fig.5(a). There is a distinct difference between Fig.3(a), Fig.4(a) and Fig.5(a). In the plot for
Au, $ln(M_{3})$ and  $ln(M_{2})$  lie on a very narrow band while for Kr
there is a large variation. This difference reflects a wider fragment mass distribution for different events with 
the same multiplicity for Kr. A similar trend with system size is seen
in percolation indicating that this is a finite size effect \cite{elliott94}.
Fig.5(b) shows the fit for Kr.  
 A nonlinear behavior is clearly observed.
This contrasts with the linear behavior seen in the corresponding
plots for Au and La in Figs.3(b) and 4(b) respectively. 
An exponential fit to the data is shown in Fig.5(b) to guide the eye. 
The fitting region was chosen by the criteria laid down in case of Au and La.
A linear fit to the Kr data gives a value of $\tau =1.88\pm0.08$ 
 with an unacceptably large $\chi^2/dof$ of 20. 
This result is consistent with Fig.1, 
in which the peak $\gamma_{2}$ value is $<$ 2 for Kr. 
\begin{figure}[ht]
\epsfxsize=8.5cm
\centerline{\epsfbox{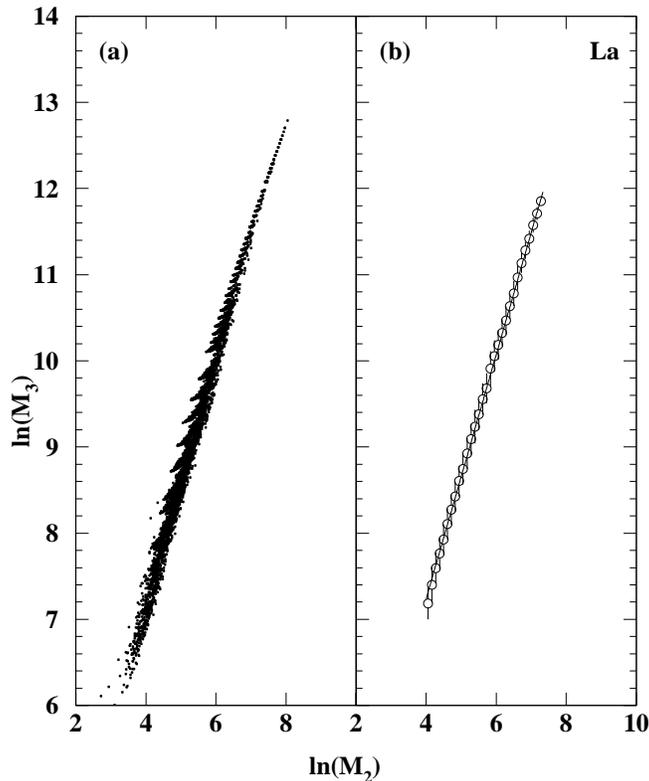}}
\caption{
a).$ln(M_{3})  vs  ln(M_{2})$ for La above the critical multiplicity.
b). The average value of $ln(M_{3})$ at a given $ln(M_{2})$.
}
\label{fig.4}
\end{figure} 

 The thermal excitation energy per nucleon at $m_{c}$, $E_{c}^*$, was 
obtained for each system from the variation of  $E^*_{th}$ with 
{\it m\/} as shown in Fig.2(b).
The dependence of  $E_{c}^*$ on system size is shown in Fig.6(a), where the size of fragmenting system is the average remnant mass at $m_{c}$ \cite{hauger00}.
The width of the remnant mass distribution at  $m_{c}$ is indicated by the 
horizontal error bars and is $\sim$ 6-8\% \cite{hauger98,hauger00}.
Fig.6(b) shows the isotope freeze-out temperature, $T_{He-DT}$, obtained from the $H^2/H^3$ and  $He^3/He^4$ double isotope
ratios at $m_{c}$\cite{hauger00,albergo}. 
Both  $E_{c}^*$ and  $T_{He-DT}$ decrease with increasing system size.
We can compare these results with calculations which have studied highly
excited nuclear matter.  The temperature-dependent Hartree-Fock(HF) calculations for 
equilibrated hot nuclei show that Coulomb repulsion causes the compound nucleus 
to become unstable at a lower temperature than the uncharged system \cite{bonche}. 
The trend seen in the present work is also seen in a HF calculation using a
Skyrme interaction with a soft equation of state\cite{levit}. This temperature
is shown in Fig.6(b) as $T_{limit}$. In another study\cite{jaqaman}  it was found that
finite size effects and Coulomb force lead to a considerable reduction
in the ``critical'' temperature. 
\begin{figure}[ht]
\epsfxsize=8.5cm
\centerline{\epsfbox{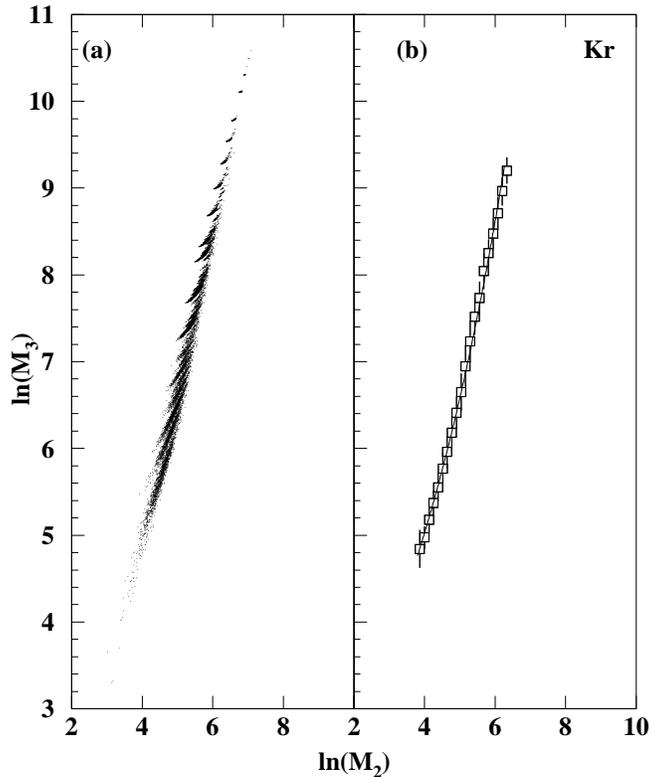}}
\caption{(a).
$ln(M_{3})  vs  ln(M_{2})$ for Kr above the critical multiplicity.
(b). Average  $ln(M_{3})$ as a function of  $ln(M_{2})$ . 
}
\label{fig.5}
\end{figure}

The Au, La and Kr results can also be compared
with the statistical multifragmentation
model (SMM)\cite{scharen,bondorf}. The SMM $E_{c}^*$ values 
are shown in Fig.6(a). 
The agreement between data and SMM is good although a $\sim$ 1 MeV/A
discrepancy is observed for Kr. 
The SMM breakup temperature $T_{SMM}$\cite{bondorf} is shown in Fig.6(b).
There is a decrease  in both temperatures with increasing system size. It is apparent 
 that the $T_{He-DT}$ temperature is about 1 MeV lower than
the SMM temperature. This difference 
is due to the fact that $T_{He-DT}$ is measured after secondary decay has taken place, 
while $T_{SMM}$ corresponds to the breakup configuration.
Particularly interesting is the fact that the 
 experimental $T_{He-DT}$ tracks 
$T_{SMM}$ in its dependence on system size at  $m_{c}$.
 SMM indicates that the decrease in  both $T_{SMM}$ and in  $E_{c}^*$ with
increasing system size is due to the increase of the Coulomb energy. 
This result
suggests that the Coulomb energy plays a central role in the MF of nuclei.
\begin{figure}[ht]
\epsfxsize=8.5cm
\centerline{\epsfbox{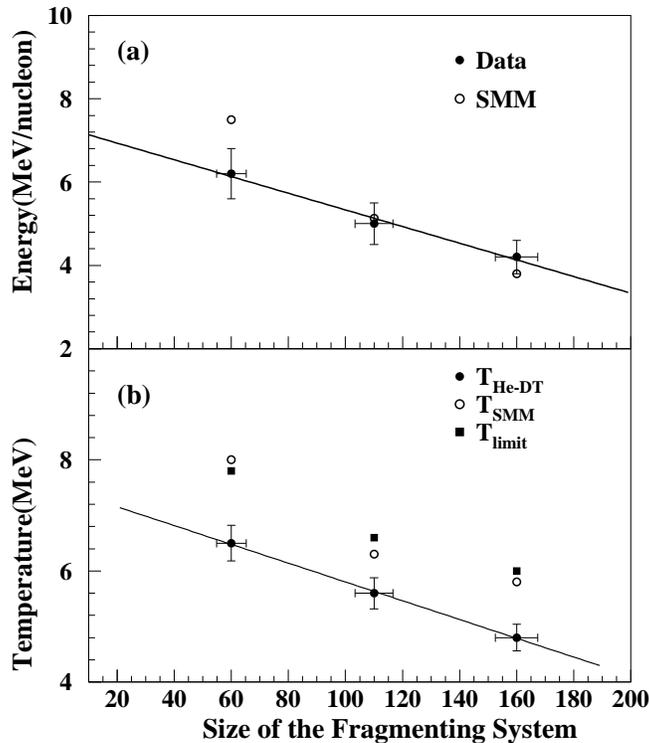}}
\caption{
a). Energy (MeV/nucleon) at $m=m_{c}$ . b). $T_{He-DT}$,
$T_{SMM}$ and $T_{limit}$ as a function of the  system size. The lines through
the points are linear fits to the data.
}
\label{fig.6}
\end{figure}  
The microcanonical Metropolis Monte Carlo (MMMC) \cite{gross1,gross2} calculations have 
emphasized that  MF is controlled by the competition between long range
Coulomb forces and finite size effects.
Finite size effects in models with only short range forces predict an 
$\it increase$ in the critical temperature as the system size increases, 
as is evident from
percolation\cite{stauffer2} and Ising model studies\cite{carmona}. Since the experimental temperature 
exhibits the opposite dependence on system size, it is apparent that
Coulomb effects are more important than finite size effects.
For finite  {\it neutral\/} matter the critical temperature ($T_{c}$)
 is expected to be $\sim$ 15-20 MeV \cite{jaqaman,lattimer}.
 The observed $T_{c}$ for A=160 is $\sim$ 6 MeV. 
Compared to finite uncharged nuclei, the presence of Coulomb energy plays
a role in lowering the excitation energy needed to reach the
regime where critical signatures are observed. In the smaller
Kr system there is less Coulomb energy in the initial remnant state.
Achieving multifragmentation in this system requires greater excitation
energy/nucleon compared to Au and La (as shown in Fig.6) and as a result, the
dynamics of the ensuing disassembly may not take this system near its
critical regime.

In conclusion, we have analyzed the fragment distributions resulting from 1A
GeV Au, La, and Kr on carbon. The reduced variance
$\gamma_{2}$ has a peak at the multiplicity where the fluctuations in 
 $A_{max}$ are largest. The peak
value of $\gamma_{2}$ is $>$ 2 for Au and La and they exhibit a power law 
fragment yield distribution at $m_{c}$.  
The peak value for Kr is  $<$ 2 and this system does not exhibit a power law
with $\tau \geq 2$. The decrease in $\gamma_{2}$ with decreasing
 system size can be attributed to finite size effects.
 These  
observations argue against a continuous phase transition in the MF of Kr but 
 are consistent with such a transition in the MF of La and Au. 
 Recent analysis based on the SMM microcanonical
caloric curve\cite{scharen}, which indicated a first order phase transition 
for the MF of Kr and a continuous phase transition for the MF of Au is
consistent with experimental observations.
The observed decrease in excitation energy and temperature with an
 increase in system size for MF at the critical point
shows the importance of the Coulomb energy in MF.

This work was supported by the U. S. Department of Energy.

\end{document}